\documentclass[useAMS,usenatbib]{mn2e}
\usepackage{epsfig}
\usepackage{times}

\newif\ifAMStwofonts
\newcommand{\uc}{ULTRACAM}
\newcommand{\sloanu}{$u'$}
\newcommand{\sloang}{$g'$} 
\newcommand{\sloanr}{$r'$} 
\newcommand{\sloani}{$i'$}
\newcommand{\target}{Swift\,J1357.2--0933}

\newcommand{\Msun}{$\rm M_{\odot}$}
\newcommand{\Rsun}{$\rm R_{\odot}$}

\newcommand{\RL}{$R_{\rm L1}$}
\newcommand{\kms}{$\rm km\,s^{-1}$}
\newcommand{\ergs}{$\rm erg\,s^{-1}\,cm^{-2}$}
\newcommand{\av}{$A_{\rm v}$}
\newcommand{\erg}{$\rm erg\,s^{-1}$}

\title[Evidence for quiescent synchrotron emission in \target] 
{Evidence for quiescent synchrotron emission in the black hole X-ray transient \target}

\author[T.\,Shahbaz et al.]
       {T.\,Shahbaz,$^{1,2}$\thanks{E-mail: tsh@iac.es}
	D.M.\,Russell$^{1,2}$, 
	C.\,Zurita$^{1,2}$,
        J.\,Casares$^{1,2}$ 
        J.M.\,Corral-Santana$^{1,2}$,
 	\newauthor 
        V.S.\,Dhillon$^3$,
        T.R.\,Marsh$^4$ \\
$^1$Instituto de Astrof\'\i{}sica de Canarias (IAC), E-38200 La Laguna, Tenerife, Spain \\
$^2$Dept. Astrof\'\i{}sica Universidad de La Laguna (ULL), E-38206 La Laguna, Tenerife, Spain \\
$^3$Department of Physics and Astronomy, University of Sheffield,
    Sheffield, S3 7RH, UK  \\
$^4$Department of Physics, University of Warwick, Coventry CV4 7AL, UK \\
}

\pagerange{\pageref{firstpage}--\pageref{lastpage}}
\pubyear{2013}

\begin{document}
\maketitle

\begin{abstract}
\noindent

We present high time-resolution \uc\ optical  and NOTCam infrared observations
of the edge-on black hole  X-ray transient \target. Our data taken in 2012 and 2013 
show the system to be at its pre-outburst magnitude and so the system is in
quiescence. In contrast to other X-ray transients, the quiescent light curves  of
\target\  do not show the secondary star's ellipsoidal modulation. The optical
light curve  is dominated by variability with an optical fractional rms of
$\sim$35 per cent, a factor of $>$3 larger than what is observed in other
systems at similar time-resolution. Optical flare events lasting 2--10\,min with amplitudes of up to 
$\sim$1.5 mag  are seen as well as  numerous rapid $\sim$0.8\,mag dip events which are similar to the optical dips seen in outburst.  
Similarly the infrared $J$-band
light curve is dominated by variability with a fractional rms of $\sim$21 per
cent and flare events lasting 10--30\,min with amplitudes of up to $\sim$1.5 mag 
are observed.

The quiescent optical
to mid-infrared spectral energy distribution in quiescence is dominated by a
non-thermal component with a power--law index of --1.4, (the broad-band rms SED
has a similar index) which  arises from optically thin synchrotron emission most
likely originating in a weak jet; 
the lack of a peak in the spectral energy distribution rules out
advection-dominated models.

Using the outburst amplitude--period relation for X-ray transients we estimate
the  quiescent magnitude of the secondary star to lie in the range
$V_{\rm min}$=22.7 to 25.6, which when combined with the absolute magnitude of the 
expected 
M4.5\,V  secondary star allows us to constrain to the distance to lie in the range 
0.5 to 6.3\,kpc. The short orbital period argues for a nuclearly evolved star 
with an initial mass $\sim$1.5\,\Msun, which has evolved to a 0.17\,\Msun star.
The high Galactic latitude of \target\ implies a
scale height in the range 0.4 to 4.8\,kpc above the Galactic plane, possibly 
placing \target\  in a
sub-class of high-$z$ short-period black hole X-ray transients in the Galactic
Halo.

\end{abstract}


%
\begin{figure*}
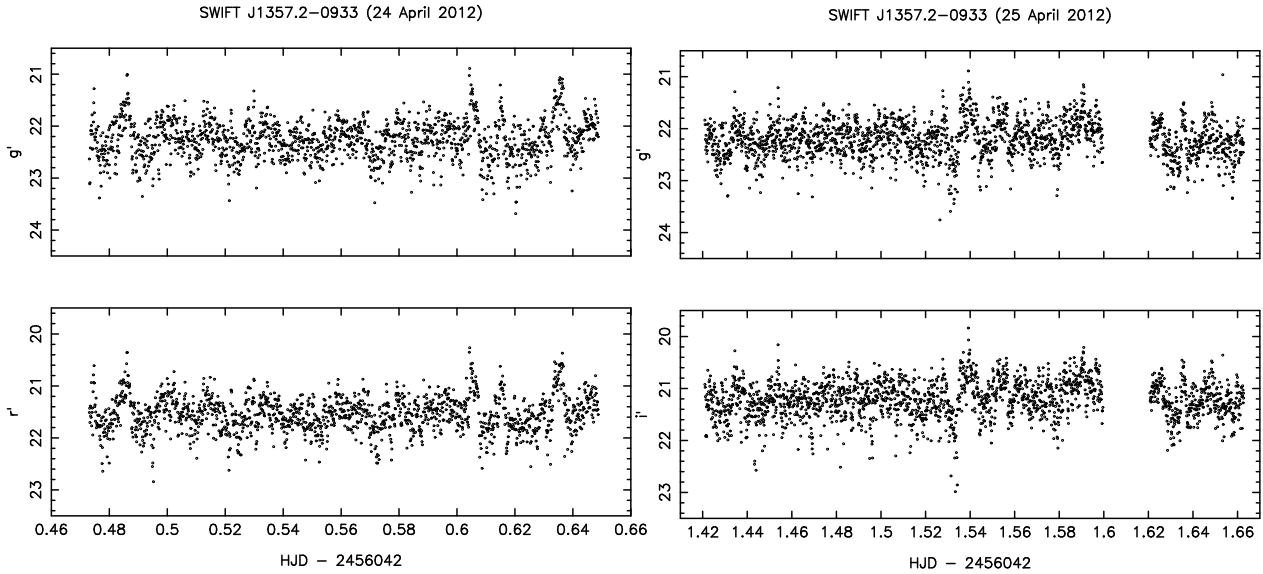

\hspace*{-85mm}
\psfig{angle=0,height=7.5cm,file=lc_24aprilP.eps}

\vspace*{-75mm}
\hspace*{80mm}
\psfig{angle=0,height=7.5cm,file=lc_25aprilP.eps}

\caption{
The observed \uc\ Sloan-band light curves of \target\ taken on UT 2012 
April 24 and 25.
}
\label{FIG:LC}
\end{figure*}
%

%
\begin{figure}
\hspace*{0mm}
\psfig{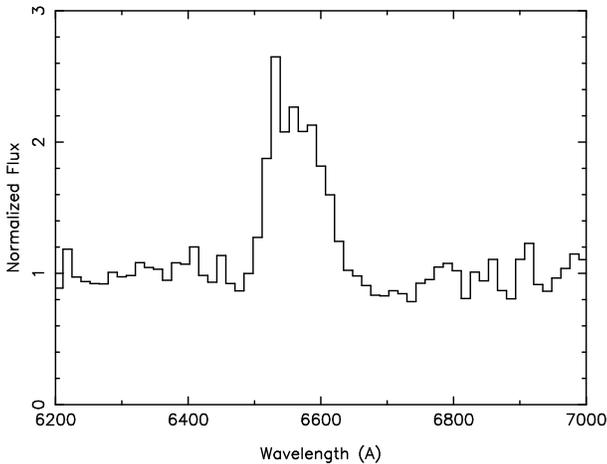}
\caption{The average normalised optical spectrum of \target\ in quiescence 
which shows strong and broad H$\alpha$ emission.}
\label{FIG:spectrum}
\end{figure}
%

%
\begin{figure}
\hspace*{0mm}
\psfig{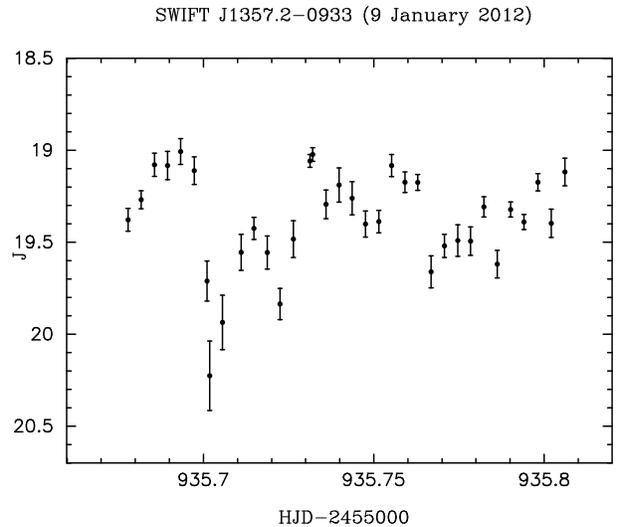}
\caption{The observed $J$-band light curve of \target\ taken on UT 2012 
January 9.}
\label{FIG:IR}
\end{figure}
%

\section{INTRODUCTION}
\label{Intro}

The new X-ray transient (XRT) \target\ with Galactic coordinates
$l$=328.702\degr\ and $b$=+50.004\degr\ was detected on 2011 January 28 by the
{\it Swift Burst Alert Telescope} \citep{Krimm11a}. \citet{Rau11} reported the
detection of the optical counterpart using the {\it Gamma-Ray Burst
Optical/Near-Infrared Detector} (GROND) taken on 2011 February 1. The images
revealed the presence of a star with $r'$=16.30 inside the X-ray  error circle.
A pre-outburst counterpart was  detected in Sloan Digital Sky Survey (SDSS)
images with $r'=21.96$\,mag, and very red colour, which suggested that the
companion star was likely a M4 star and lead to a distance estimate of
$\sim$1.5\,kpc. This strongly suggested that \target\ was a short period dwarf
nova or low-mass X-ray binary \citep{Rau11}.  The former scenario can be ruled
out though because the {\it Swift} X-ray spectrum is consistent with a pure
power--law  without a thermal component \citep{Krimm11b}. 

\citet{Torres11} reported low resolution optical spectroscopy of the counterpart
obtained early in 2011 February which showed a featureless spectrum without  any
sign of Balmer,   He\textsc{ii}\,4686\,\AA\, or Bowen emission lines, typical of
other XRTs in outburst. However,
spectroscopy taken a day later did detect a weak  H$\alpha$ emission line with 
$\rm EW\sim7$\,\AA\ and  $\rm FWZI\sim$4000\,\kms\,\citep{Milisavljevic11} and
was confirmed by \citet{Casares11} through high resolution spectroscopy which
showed a weak double-peaked extremely broad H$\alpha$ profile with velocities
characteristic of black-hole X-ray binaries (XRBs). 
From further optical spectroscopy, the motions of the H$\alpha$ emission line wings and the double-peak separation revealed a 2.8\,h orbital period and a mass function  $>$3.0\Msun, implying that the compact object in \target\ is indeed a black hole  \citep{Corral-Santana13}.

Optical light curves taken during decline from outburst showed unusual, profound
dip features that obscure up to 50 per cent of the flux, and have a recurrence
frequency which decreases during the outburst decline \citep{Corral-Santana13}.
Unexpectedly these optical dips were not present in the X-ray light curves
which, combined with the low observed X-ray luminosity \citep{Armas13} implies
that the compact object is hidden from the observer by the accretion disc as
expected for a relatively high inclination ($>$80\degr) accretion disc corona
source.

The X-ray outburst was detected for $\sim$7 months \citep{Armas13} and by  2011
August  the source was in X-ray quiescence. Here we report on our high-time
resolution multi-colour optical observations of \target\ taken during
quiescence. We report on the timing properties of the light curves  in
quiescence,  the short-term variability and the spectral energy distribution
(SED) in quiescence and near the peak of the outburst.

\section{OBSERVATIONS AND DATA REDUCTION}
\label{OBS}

\subsection{Optical photometry}

Multi-colour photometric observations of \target\ were obtained with \uc\ on
the 4.2-m William Herschel Telescope atop La Palma during the nights of UT 2012
April 24 and 25.  \uc\ is an ultra-fast, triple-beam CCD camera, where the
light is split into three  broad-band colours (blue, green and red) by two
dichroics.  The detectors are back-illuminated, thinned, E2V frame-transfer 
1024$\times$1024 CCDs  with a pixel scale of 0.3\,arcsec/pixel. Due to the
architecture of the CCDs the dead-time is essentially zero  (for further
details see \citealt{Dhillon01}).

On UT 2012 April 24 observations (UT 23:13--03:26) were taken using the
Sloan $u'$-, $g'$- and $r'$-band filters, whereas on the 2012 April 25
observations (UT 21:58--03:46) were taken using the Sloan $u'$-, $g'$-
and $i'$-band filters.  An exposure time of 10\,s was used for  the $g'$-,
$r'$- and $i'$-bands, however, given the faintness  of the target an exposure
time of 20\,s was used in the \sloanu-band.  The conditions were generally good
but the median seeing was  1.2 and 1.4\arcsec\ on 2012 April 24 and 25
respectively.

The \uc\ pipeline reduction procedures were used to debias and  flatfield  the
data. The same pipeline was also used to obtain light curves for \target\ and
several comparison stars by extracting the counts using aperture photometry. The
most  reliable results were obtained using a variable aperture which scaled with
the seeing.  The count ratio of the target with respect to a bright local
standard (which has similar colour to our target) was then determined.   The
magnitude of \target\ was then obtained using the calibrated SDSS star 
(J135716.43-093140.1) in the field. As a check of the photometry and systematics
in the reduction procedure, we also extracted light curves of a faint comparison
star similar in brightness to the target.  The mean observed magnitudes of
\target\ are given in Table\,\ref{LOG}. The quality of the \sloanu-band data is
a factor of 3 worse than the other data, which is not surprising given its
extreme faintness. Therefore in the analysis that follows  we only use the $g'$,
$r'$ and $i'$-band data, (0.1 per cent accuracy) where we can be certain that
any variability is real.

\begin{table}
\caption{Mean observed magnitudes of \target.}
\begin{center}
\begin{tabular}{lccc}\hline 
Date UT   & Band &     mean mag. & rms$^b$  \\
          &      &               &   (mJy) \\
\hline
2012 January 9    & J     & 19.38$\pm$0.28$^a$   & 6.56$\times$10$^{-3}$ \\
2012 April 24     & u'    & 23.12$\pm$0.04       &       \\
2012 April 24     & g'    & 22.26$\pm$0.38$^a$   & 2.38$\times$10$^{-3}$ \\
2012 April 24     & r'    & 21.54$\pm$0.35$^a$   & 2.06$\times$10$^{-3}$ \\
2012 April 25     & i'    & 21.21$\pm$0.36$^a$   & 1.19$\times$10$^{-3}$ \\
2013 February 3   & J     & 19.60$\pm$0.06       & \\
2013 February 3   & H     & 19.73$\pm$0.05       &  \\
2013 February 3   & K$_s$ & 18.24$\pm$0.04       & \\
2013 March 24    & J     & 19.38$\pm$0.06       & \\
2013 March 24    & H     & 19.02$\pm$0.05       &  \\
2013 March 25    & J     & 19.17$\pm$0.05       & \\
2013 March 25    & H     & 18.68$\pm$0.05       &  \\
2013 March 25    & K$_s$ & 18.14$\pm$0.05       & \\
\hline       
\end{tabular}
\end{center}
$^a$The error reflects the rms of the light curve. \\
$^b$The intrinsic source rms values.
\label{LOG}
\end{table}

\subsection{Optical spectroscopy}

We obtained three spectra on the night of 2013 April 29 using 
ACAM on the 4.2m William Herschel Telescope. The spectra covered a useful 
spectral range 5000--9000\AA\ at 6.8 \AA\,pixel$^{-1}$ and has a total exposure 
time of 80\,min (30+30+20\,min) A slit width of 1\arcsec was used which 
resulted in a spectral resolution of $\sim$700\,\kms, measured from the FWHM of the arc lines. Standard procedures were followed within \textsc{IRAF} to de-bias, flat-field  and wavelength calibrate the ACAM spectra.
The final average spectrum has a S/N of $\sim$13 per pixel in the continuum and 
is shown in Fig.\,\ref{FIG:spectrum}. 

\subsection{Infrared photometry}

Infrared (IR) $J$, $H$ and $K_s$ images of \target\  were taken with the 2.5-m
Nordic Optical telescope with the near-IR Camera and spectrograph (NOTCam) on three
different epochs. 
On UT 2012 January 9 (UT 04:23--06:59), a series of 2.6\,h of consecutive 
$J$-band 50\,s exposures were obtained where the seeing was 0.8\,arcsec. 
On UT 2013 February 3  $J$, $H$ and $K_s$-band images were taken with total 
exposure times of 27\,min (mid UT 05:25), 15\,min  (mid UT 05:51) and 15\,min 
(mid UT 06:12) respectively, with seeing ranging from 0.8 to 1.0\,arcsec. 
On the night of 2013 March 23 the conditions were cloudy. Consecutive cycles of $J$ and $H$-band images were taken with total 
exposure times of 36\,min (mid UT March 24 02:10) and 60\,min  (mid UT March 24 02:17) respectively. 
On 2013 March 24  the conditions were better with a median seeing of 0.7\,arcsec and
again consecutive 
cycles of $J$, $H$ and $K_s$-band images were taken for about 6\,h (UT March 25 00:29--06:22) with exposure times of 13.5\,min, 12\,min and 9\,min respectively. 
The resulting cycle time was 44\,min.
Basic data reduction was performed
using the NOTCam Quicklook reduction package and the relative magnitudes
determined  from \textsc{iraf/daophot} aperture photometry. Finally, we used the $J$,
$H$, and $K_s$ magnitudes of 2MASS stars the field to calibrate the photometry
and the resulting  magnitudes of \target\  are given in Table\,\ref{LOG}).
For the UT 2013 March 24 and 25 data we could only determine mean magnitudes, because of the  poor weather conditions for the former and the 
poor time resolution of 44\,min for the latter.

%
\begin{figure}
\hspace*{0mm}
\psfig{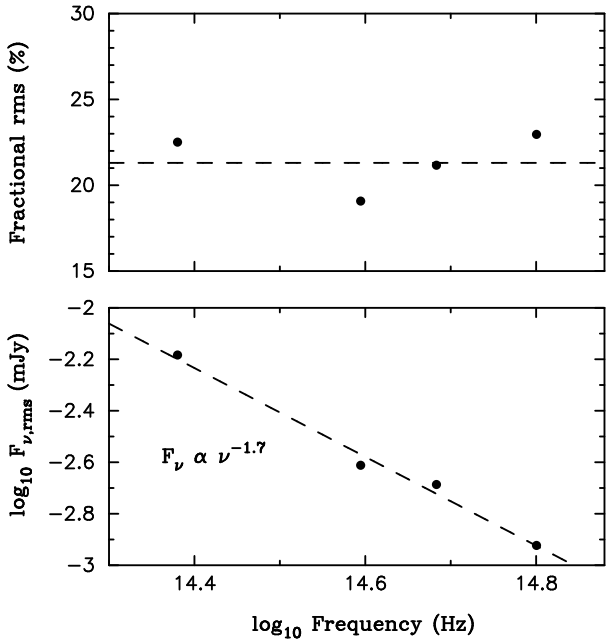}
\caption{The broad-band spectral energy distribution of the fractional rms
(top)  and rms (bottom) variability. Note that the \uc\ data has been binned 
to the same time resolution as the $J$-band data.}
\label{FIG:rms}
\end{figure}
%

%
\begin{figure}
\hspace*{0mm}
\psfig{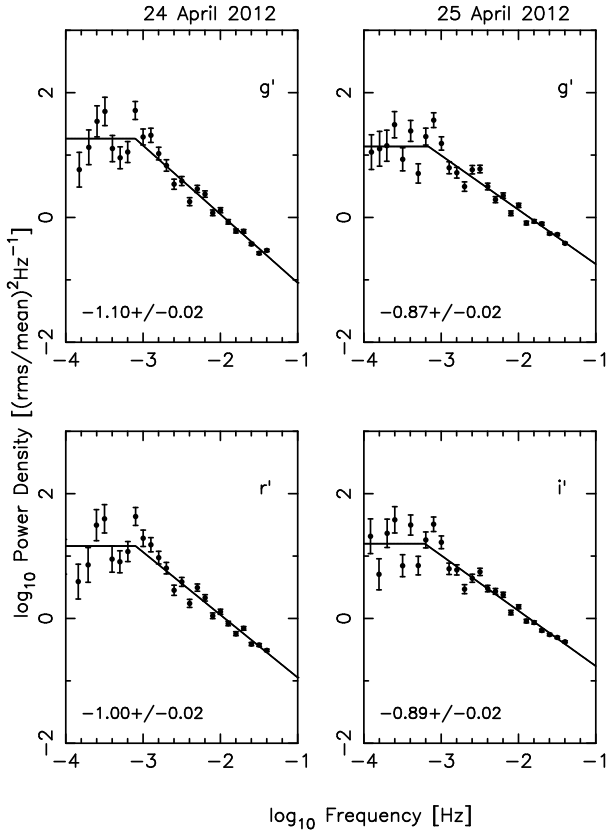}
\caption{
The power density spectra of the flare Sloan-band light curves of \target\  
taken on UT 2012 April 24 and 25. The solid line is a broken power--law  fit 
to the data and the  slope of the power-law fit is indicated  in each panel.
}
\label{FIG:PDS}
\end{figure}
%

\section{THE OPTICAL/INFRARED LIGHT CURVE}
\label{LC}

In comparison with other light curves of quiescent X-ray transients we would 
expect the optical light curve of \target\ to show the secondary star's
ellipsoidal modulation with superimposed flare events. However, as one can see
in Fig.\,\ref{FIG:LC} our Sloan-band optical light curves taken on UT 2012
April 24 and 25  only show large amplitude flares, up to $\sim$1.5 mag
superimposed on a flat light curve. There is no evidence for any orbital
modulation which suggests that another source of light with large variability
completely dominates the secondary star's optical flux. The fractional rms
variability of the optical flaring activity is large, $\sim$35 per cent in
\sloang-, \sloanr- and \sloani-bands, more than a factor of 3 larger than what
is normally observed in the optical light curves of quiescent X-ray transients with similar time resolution (\citealt{Shahbaz05}; \citealt{Shahbaz10}), and similar to the X-ray fractional
rms seen in the hard state of X-ray binaries \citep{Munoz11}. Similarly the
NOTCam $J$-band light curve taken on UT 2012 January 9 which has a  duration of
2.6\,h also shows no evidence of the secondary star's ellipsoidal modulation.
The light curve is dominated by variability with a fractional rms of $\sim$23 per cent and
flare events lasting 10--30\,min  with amplitudes of $\sim$2\,mag  are
observed (see Fig.\,\ref{FIG:IR}).

In Fig.\,\ref{FIG:LC} flare events lasting 2--10\,min are seen as well as numerous rapid
flare events which are not resolved.  There are also a number of $\sim$0.8\,mag
"dip" events observed on both nights which look similar to the dips seen in
outburst, but less frequent \citep{Corral-Santana13}.  The dips have a
characteristic duration of $\sim$2\,min and drop in brightness by up  to
$\sim$0.5--1.0\,mag, which is equivalent to a factor $\sim$2 reduction in flux,
similar to what is observed during outburst. The  recurrence period of the dips
is $\sim$30\,min, a factor of $\sim$10 longer than what is observed in outburst
\citep{Corral-Santana13}.

To see how the secondary star's ellipsoidal modulation is affected by the level
of rms variability of non-stellar flux arising from other parts of the system,
we simulated data assuming   different fractional contributions of the secondary
star light to the observed flux and different amplitudes of the rms
variability of the non-stellar light. We generated a light curve with the same sampling
as our \uc\ data. We first computed  the secondary star's ellipsoidal
modulation using the X-ray binary model described in  \citet{Shahbaz03b} and
estimates for the system parameters \citep{Corral-Santana13}. For a high
inclination, extreme mass ratio system such as  \target\ ($i$=85$^{\circ}$;
$q$=0.024) our X-ray binary model predicts a secondary star's ellipsoidal
modulation with a semi-amplitude of $\sim$0.25\,mag. To this we added a 
red-noise model light curve  generated using a power--law index of --1.0, a
break-frequency at 1\,mHz and a fractional rms variability as observed in the
data (35 per cent in the \sloang-band  light curve) calculated using the method
of \citet{Timmer95}. We then added Gaussian noise using the errors derived from
the photometry. Finally we added a constant non-stellar source  of light which
veils the fractional contribution of the secondary star. We find that even if
all the observed light arises from the secondary star (21.2 mag), 
the large rms variability of the non-stellar light component dominates over the ellipsoidal modulation.
If we decrease the fractional rms variability to 10 per cent the
secondary star's modulation can be discerned at the 4.5-$\sigma$ level, even if it only contributes 50 per cent to the observed flux (22.0 mag). In other systems in quiescence,
the red-noise and the secondary star's ellipsoidal  modulation are both observed
because  the fractional rms variability is much lower, $<$10 per cent
\citep{Shahbaz05, Shahbaz10}. 

We can compare the broad-band rms variability on the optical and IR bands and
examine the broad-band SED.  
The fractional rms of the \uc\ data is $\sim$35 per cent. However, to directly compare the 
\uc\ data which has  a time resolution  of 10\,s to the IR light
curve which has a time resolution of 300\,s, we have to bin the light curves to the same same time resolution of 300\,s (the data taken on 2013 March 24/25 was not used because of the poor time-resolution of 44\,min).
In Table\,\ref{LOG} we give the intrinsic target 
rms of the binned light curves after subtracting the noise level using stars  of
similar magnitudes in the field. The fractional rms of the \uc\ and $J$-band data 
is $\sim$21 per cent.
The resulting rms and fractional rms SEDs are shown in Fig.\,\ref{FIG:rms}.
The rms SED can be represented by a single power--law  $F_{\nu,\rm
rms}\propto\nu^{-1.7\pm0.2}$, similar to the broad-band mid-IR to optical flux SED
(see section\,\ref{SED:LOW}).

%
\begin{figure*}
\hspace*{-80mm}
\psfig{angle=-90,width=8.0cm,file=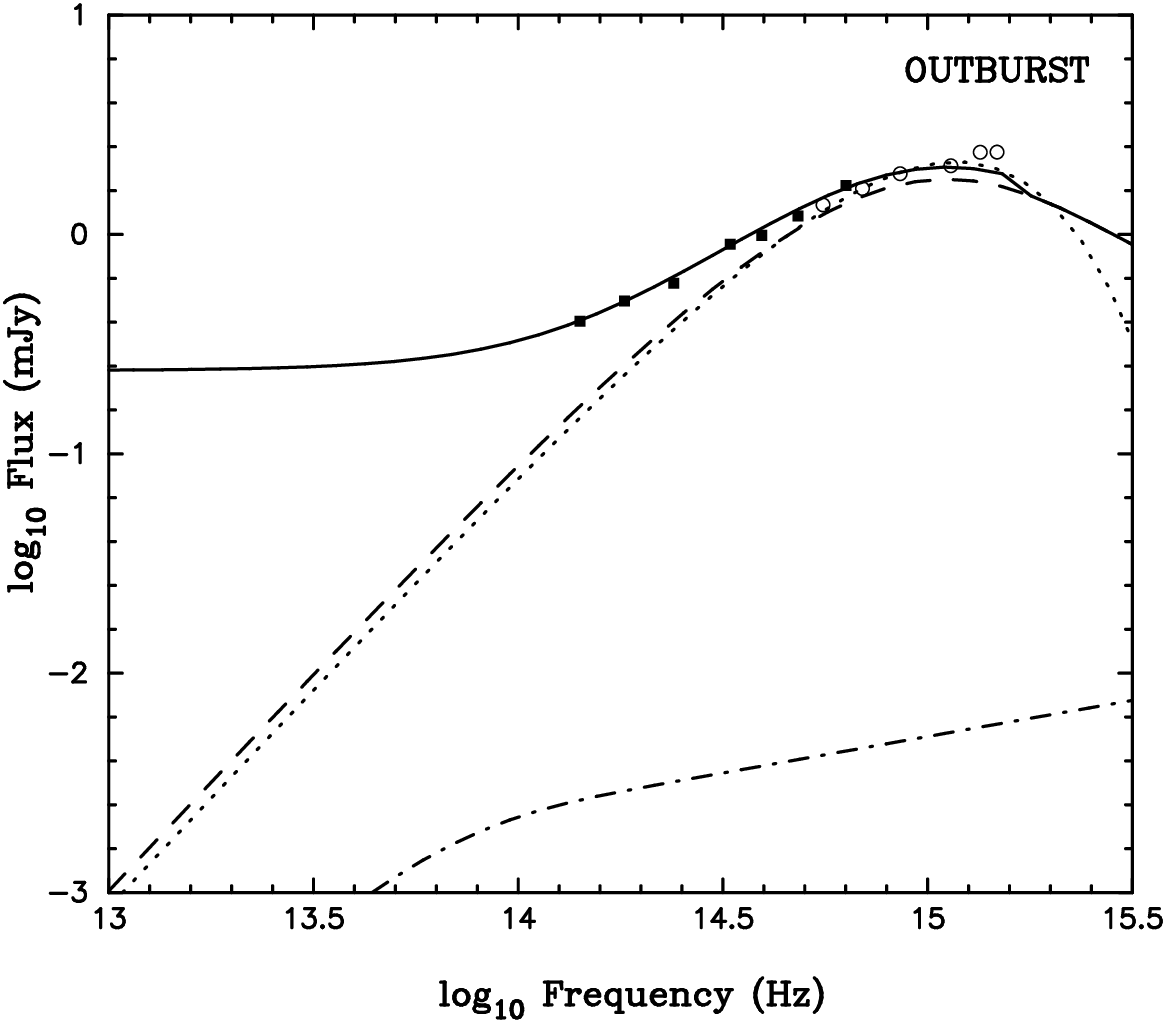}

\vspace*{-71mm}
\hspace*{85mm}
\psfig{angle=-90,width=8.0cm,file=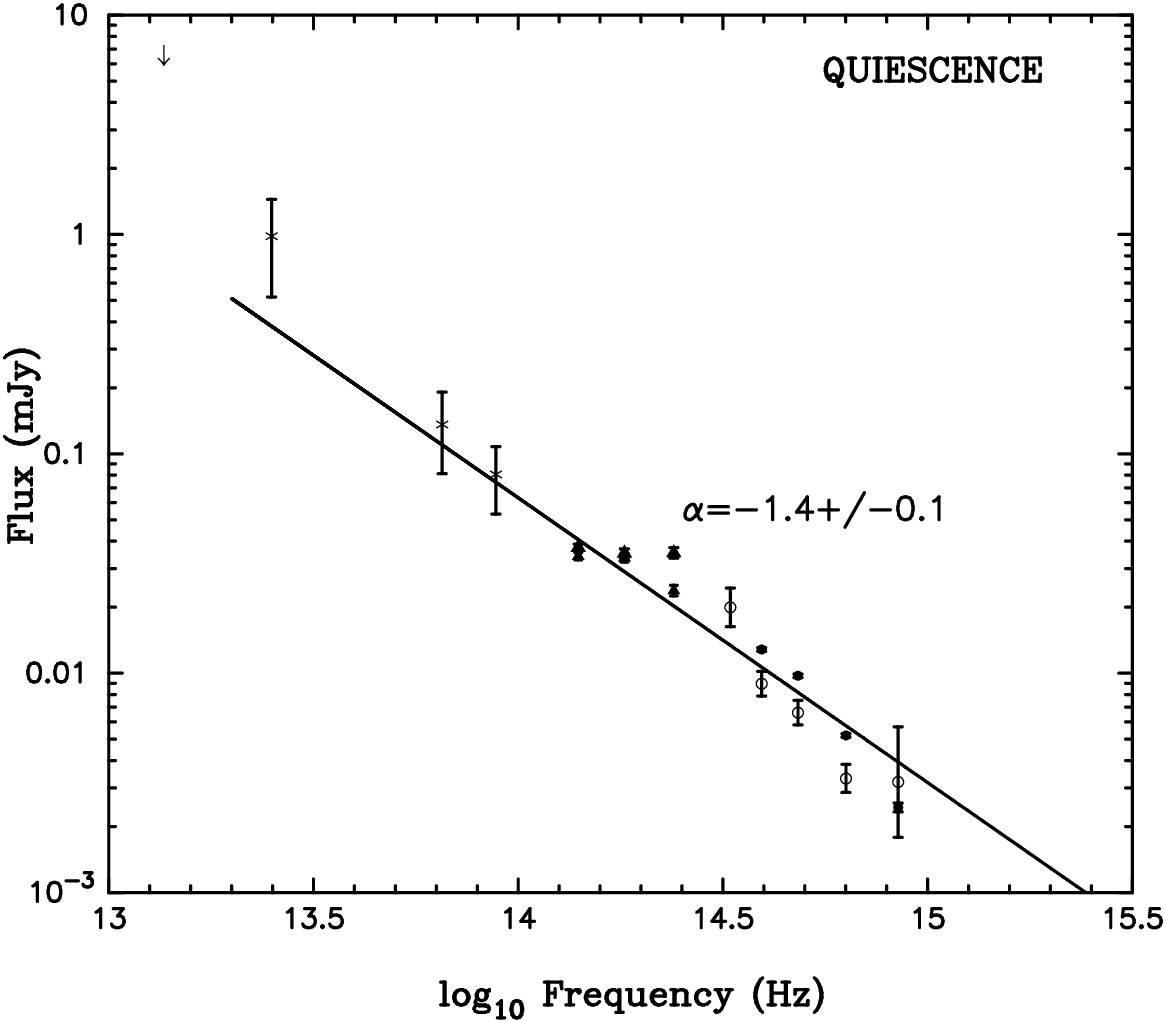}

\caption{
Left: The dereddened spectral energy distribution taken during outburst. 
The optical/UV points from Swift/UVOT
points (open circles) are taken from \citet{Armas13} and the optical/IR points from GROND (solid circles) are taken
from \citet{Rau11} on the same date. The dashed line shows an irradiated multi-colour accretion
disc model, the dot-dashed line is a standard viscous disc model and the dotted line show a single 20,000\,K\ black body. The
solid line is the sum of a power--law that extends from the radio into near-IR and
then breaks in the optical/UV and the irradiated disc model.
Right: The dereddened spectral energy distribution taken during quiescence. The
\uc\ optical data taken on 2012 April 24--25 and the NOTCam  $J$, $H$ and
$Ks$-band data taken on 2012 February 2 are shown as filled circles and triangles
respectively. The open 
circles show the SDSS\,DR7 images taken on 2006 May 26  \citep{Rau11}, and
confirms that the  source was in the same state when the independent datasets
were taken.  The stars show the WISE data (W1--W4) taken on 2010 January 21  
\citep{Wright10}.  The solid line is a power--law fit to the \uc, NOTCam and WISE
data and implies that the same mechanism produces the optical and near-IR flux  (see
section\,\ref{SED} for details). 
}
\label{FIG:SED}
\end{figure*}
%

\section{THE OPTICAL SPECTRUM}

The quiescent optical spectrum of \target\ (see Fig.\,\ref{FIG:spectrum})
shows strong and broad  $H\alpha$ emission with an equivalent width of 
$\sim$120\AA\ and a FWHM of $\sim$3900\,\kms superimposed on a flat continuum, characteristic of short period 
black hole X-ray transients in quiescence \citep[e.g.][]{Orosz95,Torres04}.
Given the high binary inclination angle one would have expected a double-peaked line profile, but we observe a single line profile. However, this may be due to the poor velocity resolution and quality of the data.
The double peak velocity separation as well as the width of the emission line profile is expected to be smaller during outburst than in quiescence because the disk expands due to viscous instabilities during outburst and hence has regions of the disc with low velocities contribute to the line profile 
\citep{Corral-Santana13}.
During outburst the FWHM $\sim$3300\,\kms \citep{Corral-Santana13}, whereas in quiescence the FWHM is  $\sim$3900\,\kms, a factor of 1.2 larger as expected.


\section{THE POWER DENSITY SPECTRUM}
\label{PDS}

To compute the  power density spectrum (PDS) of the \uc\ data we first converted the
data from dereddened magnitude to flux in mJy and used the Lomb-Scargle
method   to compute the periodograms \citep{Press92}, using  the same
normalization method as is commonly used in  X-ray astronomy, where the power
is normalized to the fractional root mean amplitude squared per hertz
\citep{Klis94}. We used the constraints imposed by the Nyquist frequency  and
the typical duration of each observation and binned and fitted the PDS in
logarithmic space \citep{Papadakis93}, where the errors in each bin are 
determined from the standard deviation of the points within each bin.  The 
white noise level was subtracted by fitting the highest frequencies with a 
white-noise (constant) plus red-noise (power--law) model. The \sloanu-band data
are of a lower quality than the data in the other filters, so here we only
compute the PDS for the  $g'$, $r'$ and $i'$-bands. We fitted the whole PDS
with a power--law and broken power--law model and found the  broken power--law
model to be significant at the $>99.99$ per cent confidence level. The break is
seen in all bands at ~$\sim$0.7\,mHz and power--law index is $\sim$-1  (see
Fig.\,\ref{FIG:PDS}).  Note that the break frequency is similar to the
time-scale on which the dips recur (see section\,\ref{LC}).

\section{SPECTRAL ENERGY DISTRIBUTIONS}
\label{SED}

In black hole XRTs, classical flat spectrum radio jets (as seen in AGN) are
commonly observed during hard X-ray states (for descriptions of X-ray states
see  \citealt{MR06, Belloni10}) when the accretion flow structure permits a
large, vertical magnetic field. In softer X-ray states jets are observed to be
quenched at radio frequencies \citep{Corbel00, Gallo03, Russell11}, which may
result from a suppression of the poloidal field by the geometrically thin disc
which exists in the soft state \citep{Meier01}. Compact jets are most commonly 
detected in the hard state and emit both through optically thick and thin
synchrotron radiation \citep{Blandford79, Falcke95}. Their emission can be 
well-modelled by a flat or weakly-inverted power--law (optically thick regime,
$F_{\nu}\propto\nu^{\alpha}$; $\alpha$=-0.1 to +0.7) from the radio regime  to
some spectral break, beyond which $\alpha$ decreases to $-1.0<\alpha<-0.4$
depending on the electron energy distribution (optically thin regime). The
location of the break occurs in the mid-IR region \citep{Russell13}; this is a
crucial piece of information as it is closely related to the physical
conditions at the base of the jet, such as the magnetic field, the base radius
of the jet, and the total energy of the electron population.

\subsection{Outburst}
\label{SED:HIGH}

In Fig.\,\ref{FIG:SED} (left panel) we show the dereddened fluxes of \target\
taken during outburst on 2011 February 1;  the ultraviolet (UV) data 
(open circles) are taken 
from \citet{Armas13} and the optical/IR data (solid circles) 
from \citet{Rau11}. We use a colour
excess of $E(B-V)$=0.04 to deredden the data determined from the the
line-of-sight extinction \citep{Armas13}. During outburst we expect the flux to
be dominated by thermal emission from the accretion disc and possibly a
non-thermal component due to a jet. 

A black body based accretion disc model has most commonly been used to fit the
IR--UV SEDs of X-ray binaries in outburst \citep{Cheng92,Frank02}. The disc is
assumed to be axisymmetric, with temperature increasing inwards with radius
$R$, having a functional form $T(R)\propto R^{-n}$. In the commonly discussed
Shakura-Sunyaev steady-state disc \citep{SS73}, $n=3/4$ and we obtain the well
known result for the observed flux $F_{\nu}\propto \nu^{3-2/n}\propto
\nu^{1/3}$ at UV/optical frequencies.  If the disc is irradiated then the
simplest assumption of a flatter temperature distribution holds, 
where the irradiation
temperature $T_{\rm irr} \propto R^{-1/2}$ drops off more slowly as a function
of radius than the temperature due to viscous heating,  $T_{\rm visc}\propto
R^{-3/4}$. We expect the inner disc to be dominated by viscous heating and the
outer by irradiation, leading to a spectrum that turns over. For an irradiated
disc the SED exhibits a much steeper flux distribution
$F_{\nu}\propto\nu^{-1}$, differing from  what we expect from a viscously
heated disc. At low frequencies one can use the Rayleigh-Jeans limit to infer 
$F_{\nu}\propto \nu^{2}$; this is independent of $n$ and so we expect the
UV--IR SED of the disc to have a blue spectrum at low frequencies, with
$F_{\nu}\propto\nu^{2}$ that turns over in the optical/UV as
$F_{\nu}\propto\nu^{1/3}$ or $F_{\nu}\propto\nu^{-1}$ depending if the disc is
irradiated. 

The dashed line in Fig.\,\ref{FIG:SED} shows an irradiated disc model; the
model parameters are for a system with an inclination angle of
$i$=80$^{\circ}$, a black hole mass  of 3.5\,\Msun\, an unabsorbed X-ray flux of
4.19$\times10^{-10}$\,\ergs\ \citep{Armas13},  a distance of 5\,kpc\ (see
section\,\ref{DIST}) and a disc radius that extends to the tidal radius of
0.9\RL (1.7\,\Rsun; \RL is the distance to the inner Lagrangian point). 
The dot-dashed line shows the standard 
viscous disc model with the same physical  parameters as the 
irradiated disc model. As one can see it substantially underestimates the observed flux.
The dotted line shows a single 20,000\,K\ black body with an arbitrary radius
of 0.47\,\Rsun and a distance of 5\,kpc. 
It is clear that there is an excess in
the near-IR that cannot be reproduced by irradiated disc model alone. 
However, if we
assume a flat SED ($\alpha$=0.0) that extends from the radio with an observed 
flux of 0.245\,mJy at 6.249\,GHz measured at the same epoch \citep{Sivakoff11}
into the IR and breaks in the optical/UV, then the resulting summed model is a
good fit to the data, as shown by the solid line in Fig.\,\ref{FIG:SED}.

\subsection{Quiescence}
\label{SED:LOW}

Pre-outburst SDSS\,DR7 images taken on 2006 May 26 show \target\ to be at
$g'$=22.8 \citep{Rau11}. Our post-outburst 
\uc\ data taken on 2012 April 24 and 25 show the system to be at $g'$=22.3 and
so given the  large  up to $\sim$ 1.5\,mag intrinsic variability observed, we can
assume that the \uc\ 2012 and SDSS\,DR7  2006  datasets were taken when the
target was in the same state. Similarly the NOTCam data taken in 2012 and 2013
have similar flux levels and so the target is in the same state. In the
WISE catalogue \citep{Wright10}, there are detections of \target\ taken on 2010
January 21  at 3.4\,$\mu$m (W1), 4.6\,$\mu$m (W2), 12\,$\mu$m (W3) and
22\,$\mu$m (W4; upper limit), a year  before the outburst. Plotting the WISE,
NOTCam and \uc\ fluxes we can see that the WISE data are a power--law extension
of the quiescent optical/IR fluxes (see Fig.\,\ref{FIG:SED}). 
Given the fact that the 2012 optical/IR data and the
2010 WISE data can be represented by the same power--law model, it is safe to
assume that the system was in the same state when these observations were taken
i.e. in quiescence. Fitting the 
\uc (4 points), NOTCam (6 points) and WISE data (3 points)  with a power--law model gives an index $\alpha$=--1.4$\pm$0.1.  We also tried fitting the data with a two-component model; power--law plus M4 star taken from \textsc{iraf/synphot}; the  radius of the M4 star was fixed at 0.29\Rsun  \citep{Corral-Santana13} and the
distance was left as a free parameter. However, an F-test shows that the two-component model is only better than the single component model at 0.05 per cent confidence level. Hence the broad-band mid-IR to UV flux SED is best described by a single power--law model.

\section{DISCUSSION}
\label{Dis}

\subsection{The quiescent spectral energy distribution}

We can compare the flares in \target\ to other quiescent black hole XRTs.  
\target\ displays the largest amplitude variability seen in any quiescent XRT
reported to date; it shows large flaring up to $\sim$1--2\,mag in amplitude.
Given the power--law index for the SED and the high inclination angle (which 
minimises any dilution from the disc)  this tells us that  any thermal
component i.e. from the secondary star or accretion disc/rim is heavily
diluted  and it is safe to assume that the  observed variability originates in
the  non-thermal emission. In the following we compare the observed quiescent
SED with what is expected from an accretion disc/flow and a jet.

\subsubsection{A jet?}
\label{JET}

Here we consider the possibility that the quiescent near-IR--optical SED  that  can be
fitted by a single power--law of index $\alpha = -1.4$  (which is similar to the 
IR-optical rms SED) could also be interpreted as emission from a jet in the system. 
For optically thin synchrotron emission the only parameter which varies the spectral
index is the particle energy distribution of the emitting electrons, $\alpha_{\rm thin}
= (1-p) / 2$. If the observed quiescent power law index of $\alpha = -1.4$ is
interpreted as optically thin synchrotron, the particle energy distribution is much
higher than usually observed in black hole XRBs and AGN, $p = 3.8$ (typical values are
$2 \leq p \leq 3$).  However, a steeper spectral index could be explained by a thermal,
possibly Maxwellian distribution of electrons in a weaker jet. In one case, the
spectral index of the jet emission in XTE\,J1550--564 was seen to vary smoothly from
$\alpha = -1.5$ when the jet was fainter, to $\alpha = -0.6$ when the jet was brighter,
and was interpreted in this way \citep{Russell10a}.  Some optically thin SEDs of
XTE\,J1118+480 also have similar steeper spectral indices \citep{Russell13}. 
Some low luminosity AGN also have steep synchrotron spectra, with $-3 < \alpha < -1$
\citep{Melia01,Markoff08,Prieto10,Fer12}. In these sources the optically thick--thin spectral break resides in
the (sub-)mm to mid-IR regime, and the optically thin synchrotron emission above the
break curves down to this steep SED. Magnetohydrodynamical numerical simulations are
able to reproduce the steep spectra as radiatively cooled synchrotron emission, where
the spectral shape is dependent on the configuration of the magnetic field in the inner regions \citep{Dibi12,Drappeau13}.
It may be the case that the faint,
quiescent spectrum of \target\ could be explained by a population of
sub-relativistic electrons in a weak outflow. This would imply that the break
in the jet spectrum between the `flat' or inverted, partially self-absorbed radio power--law
and the optically thin power--law must reside at lower frequencies than the W3
WISE bandpass, $\nu_{\rm break} < 2.5\times10^{13}$ Hz. Theoretically, this
break is expected to shift to lower frequencies at lower luminosities, if other
parameters are unchanged \citep[e.g.][]{Heinz03,Falcke04}, and this constraint
is consistent with this picture since breaks have been detected at higher
frequencies in outburst \citep[e.g.][]{Gandhi11}. Observationally, other
parameters appear to change, which is likely to produce the wide scatter
observed in the relation between jet break frequency and luminosity in black
hole XRBs \citep{Russell13}.

The reason why this outflow dominates the emission in this source and not other
quiescent black hole XRBs is likely to be related to the inclination to our
line of sight. The projected disc surface area is much smaller for this edge-on
black hole XRB, but if the inner jet is not eclipsed by the disc or the star,
and if beaming effects are negligible, the ratio of jet to disc emission will
be greater. To summarize, the observed steep power--law,
with high amplitude flickering is consistent with a variable, weak jet, and
other interpretations are most likely ruled out (see below).

\subsubsection{An accretion disc?}

\citet{Narayan97} have shown that the X-ray observations of quiescent XRTs can
be explained by a two-component accretion flow model. The geometry of the flow
consists of a hot inner advection-dominated accretion flow (ADAF) that extends
from the black hole horizon to a transition radius and a thin accretion disc
that extends from the transition radius to the  outer edge of the accretion
disc.  In  principle, interactions between the hot inner ADAF and the cool,
outer thin disc, at or near the transition radius, can be a source of optical
variability \citep{Esin97}. It has been suggested that the flares observed in
quiescent X-ray transients arise from the transition radius \citep{Zurita03,
Shahbaz03a}. Indeed, QPO and low-frequency break features in the power spectra
have been used to determine the transition radius, the transition between the
thin and advective disc (ADAF) regions \citep{Shahbaz10, Narayan96}. 

In the original ADAF models the optical flux is produced by synchrotron
emission by the hot electrons in the ADAF \citep{Narayan96}. However, as
pointed out in \citet{Shahbaz03a}, it is  difficult to explain the flare
spectrum in terms of optically thin synchrotron emission, unless the electrons
follow a much steeper power--law electron energy distribution compared to solar
flares. The SED index of the flares in the optical has now been determined  in
several systems,  in \target\ $\alpha \sim$ --1.4 (see section \ref{SED:LOW}),
A0620--00 ($\alpha \sim$ --1.6; \citealt{Shahbaz04}), GU\,Mus ($\alpha \sim$
--1.2; \citealt{Shahbaz10}) and XTE\,J1118+480 (there is no index value but the
\sloani-band  flux per unit frequency interval is larger than that in the
\sloang-band; \citealt{Shahbaz05}). These indices are much steeper  than the 
electron energy distribution in solar and stellar flares;  for solar and
stellar flares the power--law index of the electron energy  distribution $\sim$2
\citep{Crosby93}, which corresponds to a SED with an index $\alpha$=--0.5.  The
steep index is also inconsistent with the hot accretion flow model proposed by 
\citet{Veledina13}  because they predict a spectral index for this component in
the IR/optical in the range -0.5 $< \alpha <$ 0.5. Their model cannot produce
the steeper spectral index that we observe in quiescence. However, it should be
noted that in the ADAF models, there are three radiation processes that are 
important: synchrotron emission, Compton scattering, and bremsstrahlung, each 
of which produce distinct and easily recognized features in the quiescent SED.
The thermal synchrotron emission in ADAFs is invariably self-absorbed  and
produces a sharp cutoff peak, with a peak frequency that depends on the mass of
the black hole:  $\nu_s \sim 10^{15} (M/M_{\odot})^{-1/2}$\,Hz.  The
synchrotron peak is in the optical band for stellar-mass black holes.
Synchrotron emission from different radii in the flow occurs at different
frequencies and the peak emission, however, always originates close to the
black hole and reflects the properties of the accreting gas near the
Schwarzschild radius of the black hole \citep{Quataert99}. For \target\ with a
$\sim$10\,\Msun\ black hole the peak frequency would lie at $10^{14.5}$\,Hz. 
Such a peak is not seen in quiescent SED of \target.

\subsection{Comparing the outburst and quiescent properties}

The optical light curves of \target\ obtained during its 2011 outburst show
low-level flickering with a fractional rms of 7 per cent (Corral-Santana priv.
comm.) on which are superposed irregular $\sim$0.5\,mag dips that are not
present in the X-ray light curves. The dips which have a sinusoidal modulation
and a time scale of minutes  obscure up to 50 per cent of the flux and have a
recurrence frequency which decreases during the decline of the outburst. As
outlined in  \citet{Corral-Santana13} assuming that the optical dip
recurrence period reflects the Keplerian frequency of particle around a
10\,\Msun\ black hole, the 2.3\,min and 7.5\,min periods observed on 2011 March
23 and May 31  respectively would have been produced at 0.12\,\Rsun\ and 0.27\,\Rsun\
respectively \citep{Corral-Santana13}. The observed properties are explained by
the presence of an obscuring toroidal vertical structure moving outwards in the
inner disc seen at  very high inclination angle \citep{Corral-Santana13}.  The
outburst SED is thermal and can be well represented by an irradiated accretion
disc  with an IR excess possible due to a flat spectrum jet (see
section\,\ref{JET}).  

Optical ($L_{\rm V}$) and X-ray ($L_{\rm X}$)  correlations are expected 
when the optical  luminosity
originates in the viscously heated or irradiated disc as both the
X-ray and optical luminosity are linked through the mass accretion
rate. For radiatively inefficient objects such as a black hole in the hard
state  the X-ray luminosity scales as $L_{\rm X} \propto \dot{m}^{2.0}$ and so
for an accretion disc spectrum around a compact object \citep{Cheng92,Frank02}
one can determine $L_{\rm V}-L_{\rm X}$ correlations for a viscous disc.  For
typical outer disc temperatures of 8000 to 15\,000\,K the expected correlation
in  the optical is $L_{\rm V} \propto \dot{m}^{0.5}$  and so  $L_{\rm V} \propto
L_{\rm X}^{0.25}$ \citep{Russell06}. \citet{vPM94} have shown that under
simple geometric assumptions, reprocessed optical emission  $L_{\rm V}$ from
disc irradiation should be proportional to the X-ray luminosity  $L_{\rm X}$ and
scales as $L_{\rm V} \propto L_{\rm X}^{0.5}$. Indeed we confirm this using our
X-ray binary model \citep{Shahbaz03b} and find $L_{\rm V} \propto L_{\rm
X}^{0.46}$. Theoretical models of the optical emission from a jet predict 
$L_{\rm V} \propto L_{\rm X}^{0.7}$ \citep{Russell06} and observations of a
sample of low mass X-ray binaries in the hard state shows $L_{\rm V} \propto
L_{\rm X}^{0.6}$, roughly consistent with either disc reprocessing or a jet 
\citep{Russell06}. 

\citet{Armas13} find that the optical/UV and X-ray fluxes are strongly
correlated during the outburst of \target\ in the form $L_{\rm UV/Optical}
\propto L_{\rm X}^{0.2}$). By comparing the index with what is expected for a
black hole accreting via a non-irradiated disc they conclude that the outburst
and decline is dominated by a viscous disc.  However, it should be noted that
they compare the optical and X-ray fluxes during the whole outburst, and so the
index they determine most likely represents the time averaged value. The slope
of the optical--X-ray correlations during the peak of outburst seems to be
slightly steeper  than the slope of the whole outburst,  which could occur if
irradiation became more prominant near the outburst peak. Indeed our fits to
the peak of outburst SED requires an irradiated accretion disc spectrum (see
section\,\ref{SED:HIGH}).

In contrast to the outburst light curves, the quiescent light curves are 
dominated by flickering with a fractional rms of $\sim$35 per cent, and dips of
0.5--1.0\,mag are also observed, but not as frequent as in the outburst data.
If either the break in the quiescent PDS (0.7\,mHz; section\,\ref{PDS}) or the
reccurance period of the dips seen in quiescence ($\sim$30\,min;
section\,\ref{LC}) are related to the dip features, seen in outburst,  then
the break/recurrence period corresponds to a Keplerian radius of
$\sim$0.6\,\Rsun; the size of the disc is $\sim$1\,\Rsun. The quiescent SED is
non-thermal and is most likely due to synchrotron emission from a jet. In
summary, the optical outburst flux and variability are dominated by a thermal
irradiated accretion disc component and a near-IR component most likely from a
jet, whereas in quiescence the mid-IR--optical observed flux and variability are
dominated by a non-thermal jet component.

\subsection{The secondary star}
\label{STAR}

\citet{Corral-Santana13} use the orbital period and the observed
semi-empirical linear relationships for cataclysmic variable secondary stars
\citep{Knigge06}, to estimate the mass and radius of the secondary star in \target, 
which corresponds to a M4.5 main sequence star.
We can estimate the magnitude  of the secondary star using the Shahbaz \& Kuulkers 
empirical relationship between the outburst amplitude 
($\Delta V$ = $V_{\rm min}$--$V_{\rm max}$) 
and the orbital period of X-ray transients \citep{Shahbaz98},  

\begin{equation}
\Delta V = 14.36 -7.63\log P_{\rm orb}.
\end{equation}

\noindent
The relation is the consequence of the secondary star brightening faster 
than the accretion disc
as the orbital period increases and assumes that the secondary star contributes
significantly to the observed light in quiescence and the accretion disc
dominates the light during outburst.
There is scatter in the relation, which is not
expected and is most likely due to uncertainties in the fractional 
contribution of the secondary star's light, inclination effects on the disc 
emission and the peak mass transfer rates \citep{Kong12}.
Using this relationship, for an orbital
period of 2.8\,h the expected outburst amplitude is $\Delta V$ = 11. The
observed  optical magnitude during outburst maximum was $V_{\rm max}$=16.2
\citep{Rau11} which then gives an expected observed $V$-magnitude in quiescence
$V_{\rm min}$ = 27.25.   However, the Shahbaz \& Kuulkers relation does not take into account the inclination angle of the binary.
High inclination systems project a
smaller accretion disc surface area, so one has to correct the observed magnitude
at maximum by 

\begin{equation}
-2.5\log[(1.0 + 1.5\cos i)\cos i]; 
\end{equation}

\noindent
where $i$ is the inclination angle \citep{Warner87, Kuulkers13}.  For an inclination angle
of 80--89$^\circ$ the correction is 1.65--4.4\,mag. Applying this correction to 
$V_{\rm min}$ then gives an expected observed quiescent magnitude of $V_{\rm
min}$=22.7--25.6 for the secondary star.

We can compare the predicted magnitude to the observed magnitude. The observed
quiescent Sloan-band magnitudes imply a quiescent $V$-magnitude of 21.8 (using
the Sloan-band colour transformation given in \citealt{Jordi06}), which is 
0.9--3.8\,mag brighter than what we expect if the quiescent light is due to the
secondary star. There appears to be an extra component which dominates the
observed quiescent flux and this dilutes the light from the secondary star.

\subsection{The distance}
\label{DIST}

The observed light curve shows no evidence of long term variability on the
orbital period, as normally observed in the optical light curves of quiescent
X-ray transients. The large amplitude expected for a high inclination system
such as  \target\ is heavily diluted by large rms variability arising from other
regions in the system. The observed dereddened colours are not consistent with
thermal emission from the secondary star or  accretion disc and the power--law
index of --1.4 suggests non-thermal emission from a jet. This has implications
on the argument used to estimate the distance, where  \citet{Rau11} estimate a
distance of $\sim$1.5\,kpc, based on the assumption that the observed ($i$-$z$)
quiescent colours are consistent with a M4 secondary star. 

We can estimate the distance using the distance modulus of the secondary star. 
Using
the absolute magnitude of a M4.5$\pm$0.8\,V star $M_{\rm V}$=11.5--14.0
\citep{Corral-Santana13,Knigge06}, the expected magnitude of the secondary star
(see section \ref{STAR}) and \av=0.12 \citep{Armas13} gives a distance in
the range  0.5 to 6.3\,kpc.
With a galactic  latitude of 50 degrees, this 
implies a scale height 0.4--4.8\,kpc above the Galactic
plane, possibly placing \target\ in a sub-class of high-$z$ short-period black hole
XRTs in the Galactic Halo.

We can now use the new estimate of the distance to \target\ to determine the
X-ray luminosity.  \citet{Armas13} measure an unabsorbed peak X-ray flux of
$4.1\times 10^{-10}$\,\ergs\  and  upper-limits on the unabsorbed quiescent X-ray
flux 1800\,d after the outburst in  the 0.5--10\,keV energy range. Assuming a
distance of 1.5\,kpc \citep{Rau11} they obtain an outburst peak luminosity of
$L_{\rm X}=1.1\times10^{35}$\,\erg\ and a quiescent X-ray luminosity limit of
$<2\times 10^{31}$\,\erg. The low peak luminosity classifies \target\
as a  $\it very-faint$ X-ray  transient (VFXT), which have peak 
luminosities $\sim10^{34-36}$\,\erg. Our new distance estimate of 0.5--6.3\,kpc
implies an outburst peak X-ray luminosity in the range
 $1.0\times10^{34}$ to $1.4\times10^{36}$\,\erg\ and
a quiescent X-ray luminosity limit of  $<8\times10^{30}$\,\erg. Although the new
values are  up to an order of magnitude higher than previously estimated, it still
classes \target\ as a VFXT.

%
\begin{table*}
\caption{Short period black hole transients ($<5\,h$). 
}
\begin{center}
\begin{tabular}{lccccccccccccc} \hline 
Object                 & $M_{\rm X}$ & $P_{\rm orb}$  & i & d     &  z     & 
V  & $A_{\rm v}$ &
$\log L_{\rm X}$  & $\log L_{\rm OPT}$ &
$\log(L_{\rm X}/L_{\rm Edd})$    &  $L_{\rm X}/L_{\rm OPT} $ & Refs.\\ 
        & (M$_{\odot}$)  &  (hr)  & ($^\circ$)  & (kpc) &  (kpc) & (mag) & (mag) & (\erg)  & (\erg)  & (\erg) & (\erg) & \\ 
\hline 
GRO\,J0422+320         & 4.0    & 5.1 & 45     & 2.5      & 0.5       & 13.2 & 0.74 & 37.41  & 35.05 & -1.3         &  230   & (1) \\ 
XTE\,J1118+480         & 8.5    & 4.1 & 68     & 1.7      & 1.5       & 12.5 & 0.07 & 35.44  & 34.73 & -3.6         &    5   & (2)\\ 
Swift\,J1753.5$-$0127  & 10$^a$ & 3.2 &  -     & 6.0      & 1.3       & 15.9 & 1.05 & 37.31  & 34.86 & -1.8         &  282   & (3)\\
Swift\,J1357.2$-$0933  & 3.5$^b$      & 2.8 & 80--89 & 0.5--6.3 & 0.4--4.8  
& 16.2 & 0.12 & 34.00--36.15  & 32.21--34.41 & -4.7$-$-2.7         &   55  & (4)  \\ 
MAXI\,J1659$-$152      & 10$^a$ & 2.4 & 65--80 & 2.3--6.5 & 0.5--1.4  & 16.8 & 0.80 & 36.77--37.67  & 33.56--34.47 & -2.3$-$-1.4         &  1608 & (5)\\ 
\hline	 
\end{tabular}
\end{center}
\begin{flushleft}

$^a$ A mass of 10\,\Msun\ is assumed for the black hole ($M_{\rm X}$) as there is no dynamical measurement.  \newline
$^b$ Estimated using the results obtained by \citet{Corral-Santana13}. \newline
$L_{\rm X}$ is in peak X-ray luminosity the range $\sim$2--11\,keV.  \newline
$L_{\rm OPT}$ is in $V$-band peak luminosity in the range 3000--7000\,\AA. \newline
($1$) \citet{Castro93, Callanan95, Gelino03}. \newline
($2$) \citet{Yamaoka00, Uemura00, Gelino06}. \newline
($3$) \citet{ Cadolle07,Zhang07, Still05}. \newline
($4$) \citet{ Rau11, Armas13}: Section\,\ref{DIST} this paper. \newline
($5$) \citet{ Russell10b, Kong12, Kuulkers13, Kong12}.
\end{flushleft}
\label{TABLE:BHs}
\end{table*}
%


\subsection{The systemic velocity of the system}
\label{SYSTEMIC}

The double peaked H$\alpha$ profiles observed in X-ray binaries  are due to
Doppler shifts in an accretion disk around the compact object \citep{Smak81}. 
Therefore, one expects that the line velocity reflects the  motion of
the compact object. However, the complex and variable shape of  the H$\alpha$
line profile makes it difficult to measure a velocity that reliably traces the
motion of the compact object. Determining radial velocities using double-Gaussian fits to the wings of the line profile which trace the motions of the compact object 
\citep{Schneider80} should in principle give a systemic velocity ($\gamma$) 
the same as that  measured from the secondary star's radial velocity curve.

As outlined in \citet{Garcia96}  observations of  GRO\,J0422+32 and A0620--00 show that the  $\gamma$ from double-Gaussian fits to the $H\alpha$ line profile is not a robust measure of the binary systemic velocity. In GRO\,J0422+32 the H$\alpha$ systemic velocity is  142$\pm$4\,\kms\ \citep{Garcia96} which is offset compared to
the systemic velocity of the system $\gamma$=11$\pm$8\,\kms\ \citep{Harlaftis99}
obtained from the secondary star's radial velocity curve.  In A0620--00 the $\gamma$ velocity measured using double-Gaussian fitting changes from year to year, most likely because of variations in the shape of the H$\alpha$ profile; measured values for $\gamma$ are 28$\pm$6\,\kms, 1.5$\pm$0.8\,\kms\ and 0$\pm$5\,\kms\ between  1990 and 1992 \citep{Haswell90, Orosz94, Marsh94}. Therefore for \target, the $\gamma$ velocity of --150\,\kms\ determined from the wings of the H$\alpha$ emission line using double-Gaussian fitting  during outburst  \citep{Corral-Santana13}, most likely does not represent the  systemic velocity of the binary. However, this can only be confirmed by determining the systemic velocity of the binary via a radial velocity study of the secondary star.

\subsection{The evolutionary history of the system}
\label{EVOLUTION}

\begin{figure}
\hspace*{0mm}
\psfig{angle=-90,width=8.0cm,file=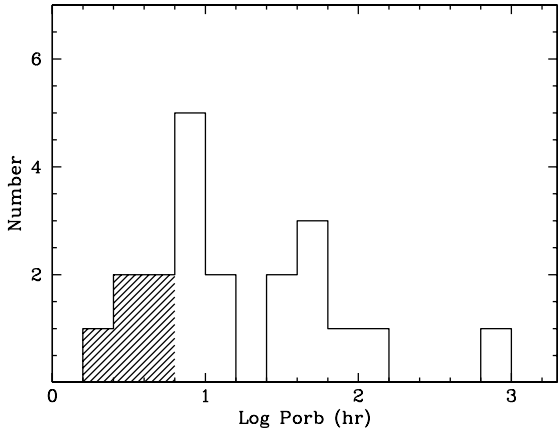}
\caption{The orbital period distribution of black hole transients.
The shaded histogram mark the group of short orbital period ($<$5\,h)
black holes systems}
\label{FIG:BH_HIST}
\end{figure}

Fig.~\ref{FIG:BH_HIST} shows the current histogram of the orbital periods of
black hole transients. In addition to the dynamical black holes the figure also
includes the  recently discovered short  period black hole transients 
Swift\,J1753.5--0127 \citep{Zurita08}, MAXI\,J1659--152 \citep{Kennea11} and
\target\ \citep{Corral-Santana13}. The distribution has clusterings at 8 and 50\,h separated by a gap. The secular evolution of low-mass X-ray binaries
follows two paths, depending  on the evolutionary stage of the companion star at
the start of the mass transfer. \citet{Pylyser88} and \citet{Ergma98} found that there is a bifurcation period  between $\sim$1\,d for the initial binary orbital period which separate converging and diverging binaries. If the orbital period
at the beginning of the mass transfer is above the bifurcation period, the
evolution of  the binary begins when the companion evolves off the main
sequence,  the mass transfer is driven by the internal evolution of the low-mass
(sub-)giant companion stars and the system will evolve towards large orbital
periods.  If the orbital period of the system at the onset of the mass transfer
is below the  bifurcation period, the companion star  is relatively unevolved,
and the only important mechanism driving mass transfer is systemic angular
momentum loss due  to magnetic braking and gravitational radiation, which
eventually leads to stripped evolved  companion stars. For angular momentum loss
due to magnetic braking the secondary star must have a convective envelope
\citep{Rappaport83}, which implies  that the initial mass of the secondary star
has to be less than 1.5\,\Msun\ (stars above this mass have radiative
envelopes). This is the same argument as outlined in \citet{Kuulkers13} for the
evolutionary state of 
MAXI\,J1659--152.  

Thus \target\ starts out with a secondary star less than 1.5\,\Msun\ and the 
evolution of the binary drives the system to shorter  orbital periods. 
\citet{Pylyser88} and \citet{Ergma98} compute the evolutionary  history of 
such types of evolution. Sequence A55 in \citet{Pylyser88}, but also see 
sequence 1 in \citet{Ergma98}, starts out with a  1.5\,\Msun\ secondary star 
mass and a 4\,\Msun\ black hole accretor, which is then evolved to an orbital 
period of 2.4\,hr, where the secondary star mass is 0.17\,\Msun, with a 
central H content of 0.09, a low mass transfer rate  $10^{-11}$\Msun/yr\ 
and an age of  5.7$\times 10^{9}$\,yr. The secondary stars in \target\ 
as well as MAXI\,J1659--152 
\citep{Kuulkers13}  and XTE\,J1118+480 \citep{Haswell02} are nuclearly 
evolved stars 
that have  undergone considerable evolution.

\subsection{Natal kicks}
\label{KICKS}

\cite{White96} determine the velocity dispersion of 
black hole X-ray binaries by studying their scale-height distribution  
and obtain a dispersion of $\sim$40\,\kms. Since the velocity dispersion of black hole  progenitors is $\sim$20\,\kms, they estimate 
an extra velocity of $\sim$20--40\,\kms\ is acquired during the formation of the black hole. This can be in the form of substantial mass-loss in  the formation of the black hole or an asymmetric natal kick. Therefore any kicks received by the black hole  affect the binary and its subsequent orbit within the Galaxy.

There is evidence for asymmetric natal kicks in the formation of neutron stars (see \citealt{Lai01} for a review). Proper motion studies of pulsars 
show that neutron stars receive natal kicks at birth in the range $\sim$200--400\,\kms\ when they are formed during the core-collapse supernova \citep{Lyne94}. Mechanisms that operate during the core-collapse that lead to the formation of a neutron star are 
large-scale density asymmetries in the pre-supernova core \citep{Burrows96} or 
asymmetric instabilites in the rapidly rotating proto-neutron star core 
\citep{Colpi02}. These can in principle operate at the time of the black hole formation, in particular if a hot proto-neutron star forms first, later 
followed by a collapse due to fall-back material. 
Peculiar velocities as large as a few 100\,\kms\ are expected for formation of 
black holes. Indeed, for the first runway black hole GRO\,J1655--50 there is 
evidence of a runaway space velocity of 112\,\kms\ \citep{Mirabel02}. 
Futhermore  \citet{Brandt95} point out that the radial velocity of the binary 
is comparable to a single neutron star having recieved a natal kick in the range 300--700\,\kms. 

XTE\,J1118+480 is located at a very high lattitude and an elevation of 
of 1.5\,kpc above the 
Galactic plane  \citep{Gelino06} and has a high space velocity of 
$\sim$145\,\kms\ relative to the local standard of rest, determined from Very Long Baseline Array measurements of the system's proper motion \citep{Mirabel01}. 
From evolutionary calculations and detailed hydronamical simulations 
\citet{Fragos09} estimate a value of the natal kick in the range 80--310\,\kms.
More recently \citet{Repetto12} have performed population synthesis calculations 
for the formatioon of black hole in X-ray binaries, including kicks due to the 
supernova mass-loss and natal kicks due to the formation of the black hole. 
They find  that large natal kicks are necessary to reach large distances above the Galactic.  However, surprising, they find that black holes  receive similar velocity natal kicks as neutron stars upon formation. 
Given the similaries of \target\ with XTE\,J1118+480 (orbital period, 
spectral type, possible large scale-height) we expect the evolution of \target\ to be 
similar (see section \ref{EVOLUTION}). 
Such short-period systems with large Galactic scale heights, could be  runaway micro-quasars that were simply kicked out of the Galactic plane into the Halo during the violent supernova explosion.

%
%

\subsection{The class of short-period black hole XRTs}
\label{CLASS}

Table \ref{TABLE:BHs} lists the current group of short period black hole
transients together with their main properties. The systems tend to display
weak X-ray outbursts with hard power--law dominated X-ray spectrum  ($\Gamma\sim
1.5-2$) and high Galactic latitudes.  Most of the  short orbital period black
hole transients are located at $z>1$\,kpc which is substantially larger than
the 0.625\,kpc dispersion of the distribution of black hole transients around the
Galactic plane \citep{Jonker04} and places them in the Halo.  Their peak X-ray 
luminosities are
also significantly lower than 0.1\,$L_{\rm Edd}$, the critical luminosity
defining the transition from the low hard state to the soft state in black hole
transients ($L_{\rm Edd}$ is the Eddington luminosity). Furthermore, the ratio
of the X-ray (2--11\,keV) to optical $V$-band luminosities is anomaly
low in XTE\,J1118+480 and  \target; X-ray binaries typically have ratios
 $\sim$100--1000, except for the accretion disc corona sources, 
which have ratios $\sim$20 (because the central X-ray source 
is hidden; \citealt{Paradijs95}).

\cite{Wu10} has shown that the peak X-ray luminosities in Eddington units
generally increase with orbital period, with no distinction between black holes
and neutron star binaries. They also noted that XTE\,J1118+480 deviates from the
trend and conclude that short orbital period $\la$4\,h black hole transients may
be radiatively inefficient  with $L_{\rm X}$ far less than 1 per cent $L_{\rm
Edd}$. Table \ref{TABLE:BHs} seems to support this conclusion which, as a matter
of fact, appears to explain why very few short period black holes have been
found thus far. Also their high Galactic latitudes  suggest a possible selection
effect since low $L_{\rm X}$ transients in the Galactic plane are more affected
by extinction and hence difficult to detect.  The short period tail in the
distribution of black hole transients seems to be finally emerging thanks to the
increased sensitivity of new X-ray satellites.

\section{CONCLUSION}

We have presented high time-resolution \uc\ optical  and NOTCam infrared observations
of the edge-on black hole  X-ray transient \target, when the source was 
at its pre-outburst magnitude i.e. in quiescence. The quiescent light curves  of
\target\  do not show the secondary star's ellipsoidal modulation and are 
dominated by variability with an optical and infrared fractional rms 
of $\sim$35 per cent, much larger than what is observed in other systems.  Optical flare events with amplitudes of up to 
$\sim$1.5 mag  are observed as well as  numerous rapid $\sim$0.8\,mag dip events which are similar to the optical dips seen in outburst. 
Similarly the infrared $J$-band
light curve shows flare events with amplitudes of up to $\sim$1.5 mag.

We find that the quiescent optical to mid-IR SED is dominated by a non-thermal component with a power--law index of --1.4, (the broad-band rms SED has a similar index), 
most likely arising from optically thin synchrotron emission from a jet, since    the lack of a peak in the spectral energy distribution rules out advection-dominated models.

Using the outburst amplitude--period relation for X-ray transients we estimate
the  quiescent magnitude of the secondary star to be in the range
$V_{\rm min}$=22.7 to 25.6. Then using the distance modulus with the expected
magnitude of a M4.5\,V star we constrain the distance to the range 0.5 to 6.3\,kpc. 
The high Galactic latitude of \target\ implies a scale 
height in the range 0.4 to 4.8\,kpc above the Galactic plane. 
This  possibly places \target\  in a sub-class of high-$z$ short-period 
black hole X-ray transients in the Galactic Halo.

\section*{ACKNOWLEDGMENTS}

We would like to thank Sera Markoff and Juan Fern{\'a}ndez-Ontiveros 
for useful discussions.
Based on scheduled, service and Director's Discretionary Time observations made with the William Herschel Telescope and the Nordic
Optical Telescope  operated on the island of La Palma by the Isaac Newton Group
and jointly by Denmark, Finland, Iceland, Norway,  and Sweden respectively,  in
the Spanish Observatorio del Roque de Los Muchachos (on the island of La Palma)
of the Instituto de Astrof\'{i}sica de Canarias.  This research has been
supported by the Spanish Ministry of Economy and Competitiveness (MINECO) under
the grant (project reference AYA2010-18080). DMR acknowledges support from a
Marie Curie Intra European Fellowship within the 7th European Community
Framework Programme under contract no. IEF 274805.
ULTRACAM, VSD and TRM are supported by the STFC.

\footnotesize{

}

\end{document}